# An explanation of infrared catastrophe of 1/$f$ power spectra

W. Chen

Simula Research Laboratory, P. O. Box. 134, 1325 Lysaker, Norway

(8 May 2003)

**Chen [1] builds the intrinsic link between the 1/$f$ power spectra and the acoustic frequency power law dissipation and, accordingly, presents two explanations of the so-called infrared catastrophe of the 1/$f$ power spectra. This note is an immediate follow-up. The major progress is to connect the 1/$f$ power spectra and the power law dissipation of an acoustic signal of finite duration, and then, we factor out the augmented function in the 1/$f$ power spectra, predicted by Mandelbrot [2]. The resolution of the infrared catastrophe puzzle is an easy byproduct of this research. In the appendix, we give a brief comment on the exponent of the 1/$f$ power spectra of intermittent turbulence.**

The 1/$f$ power spectra (fluctuation) describes that the power spectral of a signal is inversely proportional to frequency $f$ according to a 1/$f^\beta$ power law. For the power of low frequency range, for instance, $\int_0^1 P(f)df = I_0/\alpha_0 \int_0^1 1/f^\beta \, df$, $\beta \geq 1$ leads to divergence, colorfully called the infrared catastrophe. By modifying power spectra law of 1/$f^\beta$ to $f^{-\beta}R(f)$, where the augmented function $R(f)$ is assumed to vary very slowly when $f$ tends to zero, Mandelbrot [2, chapter 8] gives a reasonable explanation of this puzzling infrared catastrophe. However, $R(f)$ is a coarse conjecture without an explicit expression and Mandelbrot's explanation is more speculative, explorative, and phenomenological than theoretical. Below we try to give it an accurate and solid physical foundation through the acoustic dissipation analysis.

The dissipation of acoustic wave propagation follows

$$I = I_0 e^{-\alpha(f)t}, \tag{1}$$

where $I$ represents the power (energy, amplitude) of an acoustic signal. The attenuation coefficient $\alpha(f)$ is experimentally characterized by a frequency power law function

$$\alpha(f) = \alpha_0 f^y, \qquad y \in [0,2], \tag{2}$$

where $\alpha_0$ and $y$ are media-dependent parameters. For many media of interest, $y$ is around 1 for a broad range of frequency. The power spectra $P$ of a dissipative acoustic signal of infinite duration is given by[1]

$$P(f) = \int_0^{+\infty} I_0 e^{-\alpha_0 f^y t} dt = I_0 / \alpha_0 f^y. \tag{3}$$

(3) shows that the frequency dissipative power law (2) leads to the $1/f$ power spectra ($\beta=y$, $P \propto 1/f^y$). For a dissipative acoustic signal of finite duration, the power spectra is

$$P_T(f) = \int_0^T I_0 e^{-\alpha_0 f^y t} dt = I_0 \left(1 - e^{-\alpha_0 f^y T}\right) / \alpha_0 f^y, \tag{4}$$

where $T$ is the signal duration. Comparing (4) with (3) and, meanwhile, considering Mandelbrot's augmented function $R(f)$, one easily has

$$R(f) = I_0 \left(1 - e^{-\alpha_0 f^y T}\right) / \alpha_0. \tag{5}$$

Irrespective of the value of $y$,

$$\lim_{f \to 0} P_T(f) = I_0 T. \tag{6}$$

In terms of (6), $\int_0^1 P_T(f) df$ will not diverge when $y \geq 1$. Thus, we solve the infrared catastrophe puzzle. It is expected that most signals obeying $1/f$ power spectra in a broad range of social, economical, physical, chemical, and biological phenomena may hold the energy (not necessarily physical energy) attenuation (1) and the frequency power law (2), which lie on the solid underpinning of the corresponding fractional partial differential equation. The present puzzle resolution may thus hold in general.

[2] points out that the prefactor of the 1/*f* power spectra has something to do with the signal duration *T*. The present analysis explicitly displays this dependency in (5).

Mandelbrot [2] also analyzes the augmented function *R*(*f*) via the Wiener-Kinchin spectrum function. It will be very interesting to use the explicit expression (5) to pursue more results in this regard.

**Appendix**:

Mandelbrot [2] pointed out that intermittent turbulence dissipation may obey the 1/*f* power spectra having exponent $\beta=5/3+c$ ($c\geq 0$). In terms of the relationship between the Levy stable distribution, frequency power law dissipation, and 1/*f* power spectra (see Chen [1]), we can conclude that $0\leq y=\beta=5/3+c\leq 2$, namely, the correction $c\leq 1/3$. This finding agrees well with Mandelbrot's intuitive formula $c=(3-D)/3$, where the fractal dimension of the turbulence $D>2$ ([2], page 394).

For the generalized diffusion-wave equation (7) in Chen [1], we guess the Hurst exponent $H=\mu/s$ holds under certain conditions.